\documentclass[numberedappendix,appendixfloats,10pt]{emulateapj}
\usepackage{bm}
\usepackage{pslatex,amsmath,color}
\usepackage{graphicx}
\usepackage{natbib}
\usepackage{capt-of}
\usepackage[T1]{fontenc}
\usepackage{mathptmx}
\renewcommand{\vec}[1]{{\bf #1}}
\newcommand{\ee}[1]{{\rm e}^{#1}}
\newcommand{\vsc}{V_{\rm sc}}
\newcommand{\rms}{r.m.s.}
\newcommand{\kmin}{k_{\rm min}}
\newcommand{\kmax}{k_{\rm max}}
\newcommand{\aave}[1]{\left\langle #1\right\rangle}
\begin{document}


\title{On the interpretation of Parker Solar Probe Turbulent Signals} 

\author{Sofiane Bourouaine\altaffilmark{1} \& Jean C. Perez \altaffilmark}
\affil{Department of Aerospace, Physics and Space Science, Florida Institute of Technology, \\
  150 W University blvd, Melbourne, FL 32901, USA}

\altaffiltext{1}{email: sbourouaine@fit.edu}

\begin{abstract}
  In this letter we propose a practical methodology to interpret
future \emph{Parker Solar Probe} (\emph{PSP}) turbulent time signals even when
Taylor's hypothesis is not valid. By extending Kraichnan's sweeping
model used in hydrodynamics we derive the Eulerian spacetime
correlation function in magnetohydrodynamics (MHD) turbulence. It is
shown that in MHD, the temporal decorrelation of small-scale
fluctuations arises from a combination of hydrodynamic sweeping
induced by large-scale fluid velocity $\delta u_0$ and by the
Alfv\'enic propagation along the local magnetic field. The resulting
temporal part of the space-time correlation function is used to
determine the wavenumber range $\Delta k_\perp=[\kmin,\kmax]$ of the
turbulent fluctuations that contribute to the 
power of a given frequency $\omega$ of the time signal measured in the
spacecraft frame. Our analysis also shows that the shape of frequency
power spectrum $P_{sc}(\omega)$ of the time signal will follow the
same power-law of the reduced power spectrum $E(k_\perp)\sim 
k^{-\alpha}_\perp$ in the plasma frame, where $\alpha$ is
the spectral index. The proposed framework for the analysis of
\emph{PSP} time signals entirely relies on two simple dimensionless
parameters that can be empirically obtained from \emph{PSP}
measurements, namely, $\epsilon=\delta u_0/\sqrt 2 V_\perp$ (where
$V_\perp$ is the perpendicular velocity of \emph{PSP} seen in the
plasma frame) and the spectral index $\alpha$.

\end{abstract}

\maketitle

\vspace{0.2cm} 
\section{Introduction}
\label{sec:intro}
\vspace{0.2cm} 

The recently launched Parker Solar Probe (\emph{PSP}) mission is
expected to make in-situ measurements of the solar wind plasma from
heliocentric distances of about $r\simeq 10R_\odot$ (where $R_\odot$
is one solar radius),
near the  Alfv\'en critical point, up to distances as high as $r\simeq
200 R_\odot$ \citep{fox16}. \emph{PSP} will thus become the first
mission to explore the solar wind in the region between $r\simeq 9.5~R_\odot$ and  
$r\simeq 60R_\odot$. At these distances, the Taylor's Hypothesis
(TH)~\citep{taylor38} has been  used since the  solar wind velocity
$U_{\rm sw}$ is much higher than the propagation and turbulent
velocities of the fluctuations. This so-called frozen-in-flow TH has been
widely used to relate the power  
spectrum measured in the spacecraft frame to the reduced power
spectrum of the turbulence expected in the plasma frame using the
standard relation between the frequency of the signal, $\omega$ and
the wavenumber $k$ of the turbulent structures in the plasma frame
$\omega \simeq \vec k \cdot\vec U_{\rm SW}$~\citep[see
e.g.,][]{horbury08,alexandrova10,bourouaine12,bourouaine13,chen14}.  

As \emph{PSP} will explore the plasma of the inner heliosphere, there
has been an increased and renewed interest  in revisiting the validity
of the TH in the solar wind. Recently,
\cite{bourouaine18}, which we call BP18 hereafter, have investigated the
validity of TH   
near $r\simeq 10R_\odot$ using numerical simulations of
Reflection-driven Magnetohydrodynamic (MHD) turbulence. The authors found that
the Eulerian spacetime structure of the turbulence allows for the
interpretation of time signals even when TH is not applicable, largely
consistent with similar
works~\citep{matthaeus10,servidio11,narita13,weygand13,klein14,klein15,matthaeus16,narita17},
but with a number of important differences. For instance,
BP18 find that the Eulerian decorrelation in
simulations is consistent with 
spectral broadening associated with pure hydrodynamic sweeping by the
large-scale eddies, combined with a Doppler shift associated with
Alfv\'enic propagation along the background magnetic field. BP18, in
agreement with~\cite{narita17}, also
find that the temporal dependency of the Eulerian correlation is more
consistent with a Gaussian decay than
exponential decay found by~\cite{servidio11} and \cite{lugones16}. Another
important difference with previous works is that~BP18
find the decorrelation is the same for oppositely propagating
fluctuations even when the turbulence is imbalanced (non-zero
cross-helicity). 

In this letter, we propose a model for the Eulerian spacetime
correlation function in the context of MHD turbulence based on Kraichnan's sweeping 
hypothesis in hydrodynamics~\citep{kraichnan64}. We also show that the
proposed analytical model can be used to interpret \emph{PSP} time
signals, solely relying on two empirical parameters that can  
be easily measured from observations. 

\section{Eulerian space-time correlation }

We assume statistically homogeneous and stationary magnetized MHD
turbulence and describe the evolution of fluctuations in terms of the Elssaser variables ${\bf z}^{\pm}=\delta{{\bf {u}}}\pm\delta{{\bf v}_A}$

\begin{equation}
\frac{\partial {\bf z}^{\pm}}{\partial t} \mp {\bf v}_A\cdot{\bf \nabla} {\bf z}^{\pm} =-{\bf z}^{\mp}\cdot {\bf \nabla}  {\bf z}^{\pm}-\nabla p,
\label{eq:MHD}
\end{equation}
where ${\bf v}_A={\bf B}_0/(4\pi\rho)$ is the background Alfv\'en
velocity, $\rho$ is the density of the fluid, ${\bf \delta v}_A(\vec
x,t)$ and $\delta{{\bf {u}}}(\vec x,t)$ are the fluctuating Alfv\'en
and fluid velocity, respectively. We define the Eulerian spacetime
correlation 
function $C^\pm({\bf x},\tau)$ for ${\bf z}^{\pm}$ as
\begin{equation}
C^\pm({\bf x},\tau)=\left\langle{\bf z}^{\pm}({\bf x_0},t_0)\cdot{\bf z}^{\pm}({\bf x_0+x},t_0+\tau)\right\rangle,
\label{eq:R}
\end{equation}
where $\langle\cdots\rangle$ denotes the ensemble average over many
turbulence realizations. In the homogeneous and stationary state, the
correlation only depends on the space-time lags ${\bf x}$ and $\tau$,
and its space Fourier transform  becomes
\begin{equation}
h^\pm({\bf k},\tau)= \frac 1{(2\pi)^3}\int C^\pm({\bf x},\tau) \ee{-i {\bf k\cdot x}} d^3x,\label{eq:hdef}
\end{equation}
which is also known as the two-time energy spectrum.

We model the Eulerian correlation by extending the Kraichnan's sweeping
hypothesis (KSH), i.e., that the space-time structure of small-scales eddies 
in the Eulerian description is dominated by random sweeping by
large-scale fluctuations. In MHD, the random sweeping of
small-scale eddies by large-scale ones can occur either from the
large-scale bulk flow, which we call hydrodynamic sweeping, as well as
the wave propagation of the ${\bf z^\pm}$ along and against the local
magnetic field that results from the perturbation of the background
field by the large scale eddies, which we call Alfv\'en-wave
sweeping. This can be made evident by replacing the advecting fields
${\bf  z}^{\mp}=\delta{{\bf {u}}}\mp\delta{{\bf v}_A}$ in
Equation~\eqref{eq:MHD} to obtain
\begin{equation}
\frac{\partial {\bf z}^{\pm}}{\partial t} +\left({\delta \bf u}\mp{\bf V}_A\right)\cdot {\bf \nabla}  {\bf z}^{\pm}=0,
\label{eq:MHD2}
\end{equation}
where ${\bf V}_A={\bf v}_A+\delta{\bf v}_A$ is the \emph{local}
Alfv\'en velocity. The pressure has been ignored as its role is only
to keep the fluctuations incompressible. In equation~\eqref{eq:MHD2}
the Elsasser fields $\vec z^\pm$ undergo random advection both by the
flow $\delta\vec u$ and the local Alfv\'en velocity $\vec V_A$. We
extend KSH in MHD by replacing the advecting
variables $\delta \bf u$ and $\delta{\bf v}_A$ with zero-mean random
fields $\delta\vec u'$ and $\delta\vec v_A'$ with prescribed
statistics, which we take to be Gaussian for simplicity. Hereafter,
primed variables  indicate the field is a random variable with 
prescribed statistics. We further assume that all fluctuating fields,
$\vec z^\pm,\delta\vec{u'}$ and $\delta\vec{v}'_A$ are perpendicular
to the local mean magnetic field, namely, the direction of
$\vec{V}'_A\equiv\vec v_A+\delta\vec{v}'_A$. The space Fourier
transform of ${\bf z}^{\pm}$ then follows the linear equation
\begin{equation}
  \frac{\partial {\bf \tilde z}^{\pm}}{\partial t}  +i(\vec k_\perp\cdot\delta\vec{u'}\mp k_\|V'_A){\bf \tilde z}^{\pm}=0,
\label{eq:KraichnanMHD}
\end{equation} where $\vec{\tilde z}^\pm={\bf \tilde{z}}^{\pm}({\bf
  k},t) $ is the space Fourier transform of ${\bf z}^{\pm}({\bf x},t) $. It
is important to notice that the parallel wavenumber $k_\| $ in this
equation represents the wave-vector with respect to the local magnetic
field (along ${\bf V}'_A$) and not along the background magnetic field
(along ${\bf B}_0$).  
Equation~\eqref{eq:KraichnanMHD} is a stochastic linear equation whose
solution is 
\begin{equation}
{\bf \tilde{z}}^{\pm}({\bf k},t)  = {\bf \tilde{z}}^{\pm}({\bf k},0) \ee{\pm ik_\| V'_A t}  \ee{-i\vec k_\perp\cdot\delta\vec{u'} t }.
\label{eq:ztilde1}
\end{equation}
An important additional simplification follows for strongly magnetized
turbulence, $\delta v'_A\ll v_A$, in which case $V'_A=v_A(1+\delta
{v'_A}^2/v_A^2)^{1/2}\approx v_A$, and therefore
\begin{equation}
{\bf \tilde{z}}^{\pm}({\bf k},t)  = {\bf \tilde{z}}^{\pm}({\bf k},0)
\ee{\pm ik_\|v_A t}  \ee{-i\vec k_\perp\cdot\delta\vec{u'} t }. 
\label{eq:ztilde2}
\end{equation}
This model presents a number of significant advantages over previous
approaches based on the KSH~\citep{matthaeus10,servidio11,narita13,weygand13,narita17}. The
first is that because the random variation of $\delta\vec{v}'_A$ does
not affect the magnitude of the local Alfv\'en velocity $\vec V'_A$, to first
order in $\delta v'_A/v_A$, the Alfv\'enic sweeping is not
random. The second advantage is that in the solution provided by
Equation~\eqref{eq:ztilde2} the parallel and perpendicular components
of the wave-vector $\vec k$ are defined with respect to the direction
of the local, fluctuating magnetic field and not with respect to the
constant background field. Lastly, as we will see in more
detail later, the spectral broadening associated with sweeping solely
arises from random advection by the velocity of large-scale eddies,
and therefore affects both Elsasser components $\vec 
z^\pm$ equally. 

Assuming that $\vec{\tilde z}^\pm$ and $\delta\vec u'$ are statistically
independent at $t=0$, it is straightforward to demonstrate that the two-time power spectrum
$h({\bf k},\tau)$ defined by Equation~\eqref{eq:hdef} becomes
\begin{eqnarray}
  h^\pm({\bf k},\tau) &=&\left\langle{\bf \tilde{z}}^{\pm}({\bf k},t)\cdot {\bf \tilde{z}}^{\pm}({-\bf k},t+\tau)\right\rangle,\nonumber\\
  &=&h_0^\pm(\vec k)\ee{\mp ik_\|v_A\tau}\left\langle\ee{i\vec k_\perp\cdot\delta\vec{u'}\tau}\right\rangle,\label{eq:hsol}
\end{eqnarray}
where $h_0^\pm(\vec k)=h^\pm(\vec k,0)$ is the three dimensional power
spectrum, or the one-time ($\tau=0$) energy
spectrum. Equation~\eqref{eq:hsol} indicates that the temporal
decorrelation is the result of pure hydrodynamic sweeping,
Doppler shifted by Alfv\'enic propagation along the local
magnetic field. For simplicity we
assume that the component $\delta u_n'=\hat{\vec
n}\cdot\delta\hat{\vec u}'$ along any direction $\hat{\vec n}$ is
described by a Gaussian probability density $g(\delta\hat{u}'_n)$
where
\begin{equation}
  g(x) = \frac 1{\sqrt{2\pi}}\ee{-\frac 12 x^2},
\end{equation}
$\delta\hat u_n'\equiv \sqrt 2\delta u_n'/\delta u_0$, and $\delta
u_0=\aave{|\delta\vec u'|^2}$
is the root mean square value of $\delta\vec u'$.  Equation~\eqref{eq:hsol}
then becomes 
\begin{equation}
 h^\pm({\bf k},\tau) = h_0^\pm({\bf k})\Gamma^\pm(\vec k,\tau),
\label{eq:hspectrum}
\end{equation}
where
\begin{equation}
   \Gamma^\pm(\vec k,\tau)\equiv\ee{\mp i k_\| v_A \tau}  \ee{-\frac 14(\delta u_0k_\perp \tau)^2 }.\label{eq:Gamma}
\end{equation}
The function $\Gamma^\pm(\vec k,\tau)$ describes the temporal dependency of the two-time spectrum $h^\pm(\vec k,\tau)$ and determines the scale-dependent Eulerian decorrelation time of the turbulence.

The choice of a Gaussian probability density is made for analytical
convenience. However, the results we present here have general
validity for any other probability density, including one empirically
obtained from spacecraft data.
\section{Frequency spectrum in the spacecraft frame}

The frozen-in-flow Taylor's hypothesis is
valid in solar wind data 
when the speed of the the spacecraft seen in the plasma frame $V_{\rm
sc}=|\vec\vsc|$ is much higher than the propagation velocity $v_{\rm
ph}$ and velocity amplitudes $\delta u_0$ of the turbulent
fluctuations, and thus the frequency $\omega$ of the signal can be
related to turbulent fluctuation scale $1/k$ as $\omega\simeq|\vec
k\cdot\vec V_{\rm sc}|$. However, in our analysis we will show that
there are other cases in which we can still connect $\omega$ to $k$
even if $V_{\rm sc}\sim v_{ph}$. The key quantity that determines this
criterion is the decorrelation function $\Gamma^\pm({\bf k},\tau)$
defined in Equation~\eqref{eq:Gamma}. 

Following \cite{horbury08,bourouaine13} the power spectrum from
single-point measurements in
the spacecraft frame $P_{\rm sc}^\pm(\omega)$ is related to the three
dimensional power measured in the plasma frame by expression 
\begin{equation}
P^\pm_{\rm sc}(\omega)=\frac 1{2\pi} \int h^\pm({\bf k}, \tau) \ee{i(\omega+{\bf k}\cdot {\bf V_{\rm sc}})\tau}  d\tau d^3k,
\end{equation}
which upon substitution of $h^\pm({\bf k}, \tau)$ from
Equation~\eqref{eq:hspectrum} gives
 \begin{eqnarray}
   P^\pm_{\rm sc}(\omega)&=&\frac 1{2\pi}\int h_0^\pm({\bf k})\Gamma^\pm(\vec k,\tau) \ee{i(\omega+{\bf k}\cdot {\bf V_{\rm sc}})\tau}  d\tau d^3k,\nonumber\\
                         &=&\int h_0^\pm({\bf k})\tilde\Gamma^\pm({\bf k},\omega)  d^3k,
\label{eq:Psc1}
\end{eqnarray}
where
 \begin{equation}
\tilde\Gamma^\pm({\bf k},\omega) = \frac{1}{2\sqrt{\pi}\gamma}\ee{-\frac{(\omega\mp k_\| v_A+{\bf k} \cdot {\bf V}_{\rm sc})^2}{4\gamma^2}}.
\label{eq:Fkw}
\end{equation}
Here $\gamma= k_\perp\delta u_0/2$ represents the spectral broadening
around the Doppler shifted frequency $\vec k\cdot\vec\vsc$, the
same for both $\vec z^\pm$. 


Intuitively, the TH relies on the
assumption that the spacecraft is moving through the plasma (or the
plasma passing by the spacecraft) so fast that the
turbulence is ``frozen-in'', or simply, the turbulence does not have 
sufficient time to evolve during the observation time. The
decorrelation function contains two independent characteristic
velocities, the Alfv\'en speed $v_A$ and the velocity \rms~$\delta
u_0$, associated with Alfv\'en-wave advection and random hydrodynamic
sweeping. One can parametrize the decorrelation function with
$\epsilon\equiv\delta u_0/\sqrt 2V_{\rm sc}$ by normalizing all
velocities to $V_{\rm sc}$ and obtain 
\begin{equation}
\tilde\Gamma_\epsilon^\pm({\bf k},\omega)=\frac 1{\epsilon k_\perp V_{\rm sc}}g\left(\frac{\omega\mp k_\|v_A+\vec k\cdot\vec V_{\rm sc}}{\epsilon k_\perp V_{\rm sc}}\right),
\label{eq:Fkw2}
\end{equation}
which upon substitution in Equation~\eqref{eq:Psc1} leads to
\begin{equation}
  P_{\rm sc}^\pm(\omega) = \int h_0^\pm(\vec k)\tilde\Gamma^\pm_\epsilon(\vec k,\omega)d^3k.\label{eq:Psc2}
\end{equation}
In the limit $\epsilon\rightarrow 0$ one obtains
\begin{equation}
  \lim_{\epsilon\rightarrow 0}\int h_0^\pm(\vec k)\tilde\Gamma_\epsilon^\pm({\bf k},\omega) = \int h_0^\pm(\vec k)\delta(\omega\mp k_\|v_A+\vec k\cdot\vec V_{\rm sc}),
\end{equation}
which for existing solar wind observations $\vec\vsc\simeq
-\vec U_{\rm SW}$, with $U_{\rm SW}\gg
v_{A}$ one recovers the commonly used TH condition  
\begin{equation}
  P^\pm_{\rm sc}(\omega)=\int h_0^\pm(\vec k)\delta(\omega-\vec
  k\cdot\vec U_{\rm SW}).\label{eq:TH} 
\end{equation}
In this sense, when either
one of the two conditions $\epsilon\ll 1$ and $v_A\ll V_{\rm sc}$ no
longer hold, Equation~\eqref{eq:Psc2} should be used in lieu of the
TH. One should also note that the TH given by Equation~\eqref{eq:TH}
also holds when $v_A\sim V_{\rm sc}$ provided the turbulence is
strongly anisotropic (i.e., $k_\|\ll k_\perp$).

It is worth mentioning that the resulting model for $\tilde\Gamma_\epsilon^\pm({\bf
  k},\omega)$ only relies on the validity of the KSH, and it is not specific to a turbulence
model. Equation~\eqref{eq:Psc2} allows us in
general to relate temporal signals in the spacecraft frame to
the spatial properties of the turbulence in the plasma frame, and reduce in
the proper limits to the TH. In this sense, as we
show in this paper, these equations allow us to analyze spacecraft
signals when the TH is not valid, with the only
requirement that the KSH holds.   In the following we
proceed to explore the usefulness of the more general
Equation~\eqref{eq:Psc2} in the analysis of solar wind observations,
with focus on the upcoming measurements from the \emph{PSP}
mission.

Let us define the reduced perpendicular power spectrum
$E^\pm(k_\perp)=2\pi\int h_0^\pm(k_\perp, k_\|) k_\perp dk_\|$, and
make the following assumptions: 1) the three dimensional power
spectrum is nearly isotropic in the perpendicular plane, 2) the
spacecraft velocity in the Sun's frame, $\vec V_\perp$, is nearly
perpendicular to the magnetic field and 3) the power spectrum is highly
anisotropic, that is, nearly zero for $k_\|\ll k_\perp$. Then
Equation~\eqref{eq:Psc2} becomes    
 \begin{equation}
   P_{\rm sc}^\pm(\omega)=\int_0^\infty E_{\rm sc}^\pm(k_\perp,\omega)dk_\perp,
\label{eq:Psc3}
\end{equation}
where
\begin{equation}
  E_{\rm sc}^\pm(k_\perp,\omega)=\frac 1{k_\perp V_\perp}
  E^\pm(k_\perp)\bar g_\epsilon(\omega/k_\perp V_\perp)\label{eq:ekf}
\end{equation}
 is the spectral  density describing the energy distribution among
 frequencies and perpendicular wavenumber in the spacecraft
 frame, and the  function 
\begin{equation}
\bar g_\epsilon(x)=\frac 2{\pi}\int_0^{\pi} \frac 1\epsilon
g\left(\frac{x+\cos\phi}{\epsilon}\right) d\phi, 
\label{eq:gbar}
\end{equation}
is the average of $\tilde\Gamma_\epsilon^\pm(\vec k,\omega)$ over the angle $\phi$  between
$\vec k_\perp$ and $\vec V_\perp$.  An additional factor of two has
been added to include the contribution to $P_{\rm sc}(\omega)$ from
negative frequencies so we can assume $\omega\ge0$
hereafter. Equations~\eqref{eq:Psc3}~and~\eqref{eq:ekf} will form the
basis of our proposed methodology.

A few important aspects of the function $\bar
g_\epsilon(x)$ are worth 
emphasizing: 1) its integral
from $x=0$ to $\infty$ is equal to one, 2) it 
is the same for both $E^\pm$ energy spectra, and 3)
it is smooth for finite $\epsilon$ but its derivative becomes singular
at $x=1$ in 
the limit $\epsilon\rightarrow 0$. This last property leads to
a spectral density highly localized along $\omega=k_\perp V_\perp$
corresponding to the frozen-in-flow TH, which means that the 
energy in a small frequency band  $d\omega$ around $\omega$ entirely
arises from fluctuations with wavenumbers in the range $dk_\perp$
around $k_\perp$, with  $k_\perp=\omega/ V_\perp$.

For finite $\epsilon$, the function $\bar g_\epsilon(x)$ broadens
around $x\simeq1$ and as a result, the energy in the frequency range
$d\omega$ around $\omega$ results from a broader range of
wavenumbers, and therefore a one-to-one association between frequency and
wavenumber no longer seems possible. In fact, Equation~\eqref{eq:Psc3}
shows that the fluctuation energy in the range $d\omega$
around $\omega$ results from a non-trivial integral over a broad range
of wavenumbers weighted by $\bar g_\epsilon(\omega/k_\perp V_\perp)$.  

Let us now determine the power spectrum $P_{\rm sc}(\omega)$ when the
spatial power spectrum in the plasma frame follows a power law of the
form $E(k_\perp)=Ck^{-\alpha}_\perp$. Note that we no longer
distinguish between $E^\pm$ as the following analysis is identical for
both spectra. After changing the $k_\perp$ integration in terms of the
new variable $x=\omega/k_\perp V_\perp$ Equation~\eqref{eq:Psc3}
becomes  
 \begin{equation}
P_{\rm sc}(\omega)= \frac C{V_\perp}\left(\frac\omega{V_\perp}\right)^{-\alpha}\int_0^\infty f_\epsilon(\alpha,x)dx,
   \label{eq:Psc4}
\end{equation}
where
\begin{equation}
  f_\epsilon(\alpha,x)\equiv x^{\alpha-1}\bar g_\epsilon(x).
\end{equation}
One must note that~\eqref{eq:Psc4} is valid if the power law for
$E(k_\perp)$ extends from $k_\perp=0$ to $\infty$. 
From this result we infer the following conclusions: 1) $P_{\rm
  sc}(\omega)$ is also a power-law with the same spectral index of the
spectrum $E(k_\perp)$, consistent with the 
findings of \cite{narita17} and \cite{bourouaine18}; 2) the overall
frequency power 
spectrum is scaled, compared with the case when the
TH is valid, by a factor that solely depends on the distribution of
large-scale eddies.

Equation~\eqref{eq:Psc4} relating the power spectrum $P^\pm_{\rm
sc}(\omega)$ in the spacecraft frame to the reduced energy spectrum
can be used to define the range of wavenumbers $\Delta
k_\perp=[\kmin,\kmax]$ that
provide most of the energy at given frequency $\omega$. The mapping
between a given frequency and the range of wavenumbers providing most
of its energy, $\omega\rightarrow\Delta k_\perp$, solely depends on the
function $f_\epsilon(\alpha,x)$, determined from parameters $\epsilon$
and $\alpha$, whose values can be obtained from spacecraft
observations.

For a fixed set of values $\epsilon,\alpha$, let us define $x_{\rm
  min}$ and $x_{\rm max}$ so that  
\begin{equation}
\int_{x_{\rm min}}^{x_{\rm max}} f_\epsilon(\alpha,x) dx =\eta\int_{0}^\infty f_\epsilon(\alpha,x) dx.
\end{equation}
where $\eta$ is a dimensionless number smaller than one, representing
the desired fraction of the total energy contained between $x_{\rm
  min}$ and $x_{\rm max}$. For instance, one can choose $\eta\gtrsim
0.9$ to capture 90\% of the total energy.  We can then use
$x_{\rm min}$ and $x_{\rm max}$ to determine the wavenumber range
$\Delta k_\perp=[\kmin,\kmax]$ with
the largest contribution to a given frequency $\omega$, as
$\kmin=\omega/x_{\rm max}V_\perp$ and $\kmax=\omega/x_{\rm 
  min}V_\perp$, providing most of the power $P_{\rm sc}(\omega)$
at frequency $\omega$.

In the next section we will estimate the frequency-dependent
broadening $\Delta k_\perp$ for two different sets of
$\epsilon,\alpha$ that are representative of the regions that
\emph{PSP} spacecraft is expected to explore.

\begin{figure}[t] 
\centerline{\includegraphics[scale=0.25]{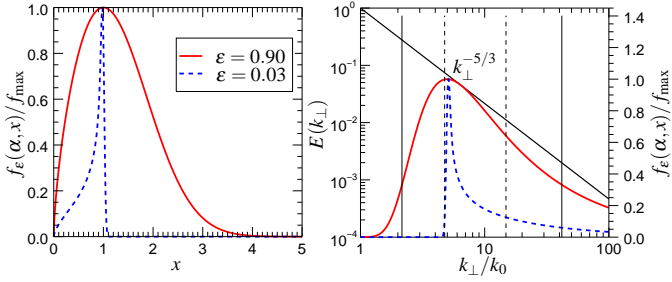}}
\caption{Left panel: Function $f(x)$ vs $x$ for $\epsilon=0.9$ close to the Sun (solid red) and $\epsilon=0.03$ near 1~AU (dashed blue). $f(x)$ is normalized to its local maximum $f_{\rm max}$. Right Panel: Hypothetical power-law energy spectrum $E(k_\perp)\propto k_\perp^{-\alpha}$ (left vertical axis) and $f_\epsilon$ vs $k_\perp$ for the same values of $\epsilon$.
  \label{fig1}}
  \vspace{0.1cm}
\end{figure}


\section{Application to \emph{PSP} data}

At \emph{PSP}'s smallest perihelion, approximately at $9.86R_\odot$,
the spacecraft velocity in the Sun's frame will be approximately
$V_\perp\sim 200~$km/s and nearly perpendicular to the magnetic
field. In the plasma frame, the spacecraft velocity is $\vec \vsc=\vec 
V_\perp-\vec U_{\rm SW}$, where $\vec U_{\rm SW}$ is the radial solar wind
velocity. \emph{PSP}'s perihelion occurs near the Alfv\'en critical
point where $U_{\rm SW}\sim v_A$, therefore, based on our strong
anisotropy assumption $\vec k\cdot\vec\vsc\simeq\vec
k_\perp\cdot\vec V_\perp$. We assume that the \rms~of velocity
fluctuations at this heliocentric radius is $\delta 
u_0\simeq250~{\rm km/s}$, which should decrease above 
the Alfv\'en critical point $r\simeq 10R_\odot $ according to turbulence
models \citep{cranmer12,perez13}. As a consequence, the parameter
$\epsilon=\delta u_0/\sqrt{2}V_\perp$ is expected to decrease with
increasing heliocentric distance $r$, where its highest value is $\epsilon\simeq0.9$   
at $r=10R_\odot $ and its lowest value is about 0.03 near 1~AU.
 
Assuming a spectral index $\alpha=5/3$ (Kolmogorov turbulence), we can
construct the function $f_\epsilon(\alpha,x)$ versus $x$ for
representative values $\epsilon\simeq0.03$ near 1~AU and 0.9 near
\emph{PSP} perihelion (left panel of Figure 1).  It can be seen that
$f_\epsilon(\alpha,x)$ is relatively narrow around $x\simeq1$ for the
small value of $\epsilon$, while for $\epsilon\simeq0.9$ significant
broadening occurs for $f_\epsilon(\alpha,x)$ around its peak value,
which is close but not equal to one. Therefore, we anticipate that
$f_\epsilon(\alpha,x)$ will be much broader near the Alfv\'en critical
point than around 1~AU.  The right panel of Figure 1 shows a
hypothetical power law spectrum (on the left vertical axis) vs
$k_\perp/k_0$ spanning
two decades, where $k_0$ is some characteristic wavenumber. Just below
the power law spectrum, the  function
$f_\epsilon(\alpha,\omega/k_\perp V_\perp)$ is shown (on the right
vertical axis) vs $k_\perp/k_0$ for a selected frequency
$\omega=5k_0V_\perp$. The two plots corresponding to the same values
of $\epsilon$ in the left panel show the contrast in the
interpretation of the same power law at a given frequency. The
vertical bars indicate the range of wavenumbers that contribute to
about $90\%$ of the energy. Near Earth's orbit, most of the energy at
each frequency is sharply localized around $\omega\simeq k_\perp
V_\perp$, whereas the same amount of energy is spread over a wider
range of wavenumbers near the Sun.

\section{Conclusions}
In this letter we introduced an analytical model for the two-time energy
spectrum given by Eq.  \eqref{eq:hspectrum} based on  two minimal
assumptions that apply to a wide range of solar wind conditions:  
1) the temporal decorrelation for the Eulerian fields $\vec z^\pm(\vec
x,t)$ is a consequence of random sweeping of the small-scale
eddies by large-scale ones; and 2) the turbulence is strongly
magnetized $\delta v'_A\ll v_A$. It then follows that the
decorrelation in time of the turbulent eddies is controlled by
random sweeping due to large-scale fluid velocities and by
pure Alfv\'enic propagation. This seems to be consistent with earlier
obtained results using numerical simulations of strongly MHD
turbulence \cite{lugones16,bourouaine18}. 
 
The analytical model for the two-time energy spectrum was used to
develop a methodology to connect time signals to the spatial
properties of the underlying solar wind turbulence under typical
conditions that \emph{PSP} might encounter. The proposed method solely
depends on the two measurable parameters, $\epsilon$ and $\alpha$,
from where one can determine $x_{\rm min}$ and $x_{\rm max}$ to estimate the
broadening in $k_\perp$ as $\Delta k_\perp=[\kmin,\kmax]$ for a
given frequency $\omega$ such that, $\kmin=\omega/x_{\rm max} V_\perp$ and
$\kmax=\omega/x_{\rm min} V_\perp$. For example, the right panel of
Figure~\ref{fig1} shows a hypothetical Kolmogorov power-law spectrum
$E(k_\perp)\simeq  k^{-5/3}_\perp$ together with the range of
wavenumbers that contribute to $90\%$ of the energy at a given
frequency for two different values of $\epsilon$. These parameters
were chosen to represent typical values 
expected near \emph{PSP} perihelion and 
near $1~$AU. 

The model we proposed for the two-time energy spectrum and the
resulting methodology differs from previous works in that it requires
no assumptions about the turbulence dynamics and is based on just two
parameters that can be easily calculated from data. A key physical
difference of our model with~\cite{narita17} is that the spectral
broadening is the same for both Elsasser fluctuations, as it only
results from hydrodynamic sweeping. The random variation of the
magnetic field associated with the large scale eddies only plays a
role in defining the direction of the local magnetic field along which
small eddies propagate, but it does not enter in the sweeping to first
order in $\delta v_A/v_A$. The proposed methodology also applies to
any spacecraft, including those flying in the Magneosheath (like
MMS).  Although our model was obtained for Alfv\'enic
fluctuations, we conjecture that the KSH may in principle be extended
to turbulence in kinetic scales whenever large-scale sweeping
dominates any kinetic  decorrelation timescales. More intuitively, the
KSH can be seen as the TH applied to an ensemble of systems in which
frozen small-scale structures are swept by a constant but random flow.  
However, because this regime requires a kinetic
description of the turbulence dynamics, it requires further
investigation.


\acknowledgements
This work was supported by grant NNX16AH92G from NASA's Living with a
Star Program. High-performance-computing resources were provided
by the Argonne Leadership Computing Facility (ALCF) at Argonne
National Laboratory, which is supported by the Office of Science of
the U.S. Department of Energy under contract DE-AC02-06CH11357. The
ALCF resources were granted under the INCITE program between 2012 and
2014. High-performance computing resources were also provided by the   
Texas Advanced Computing Center (TACC) at The University
of Texas at Austin, under the NSF-XSEDE Project TG-ATM100031.


\begin{thebibliography}{20}
\expandafter\ifx\csname natexlab\endcsname\relax\def\natexlab#1{#1}\fi

\bibitem[{Alexandrova {et~al.}(2010)Alexandrova, Saur, Lacombe, Mangeney,
  Schwartz, Mitchell, Grappin, \& Robert}]{alexandrova10}
Alexandrova, O., Saur, J., Lacombe, C., Mangeney, A., Schwartz, S.~J.,
  Mitchell, J., Grappin, R., \& Robert, P. 2010, Twelfth International Solar
  Wind Conference, 1216, 144

\bibitem[{Bourouaine {et~al.}(2012)Bourouaine, Alexandrova, Marsch, \&
  Maksimovic}]{bourouaine12}
Bourouaine, S., Alexandrova, O., Marsch, E., \& Maksimovic, M. 2012, The
  Astrophysical Journal, 749, 102

\bibitem[{Bourouaine \& Chandran(2013)}]{bourouaine13}
Bourouaine, S., \& Chandran, B. D.~G. 2013, The Astrophysical Journal, 774, 96

\bibitem[{Bourouaine \& Perez(2018)}]{bourouaine18}
Bourouaine, S., \& Perez, J.~C. 2018, ApJL, 858, L20

\bibitem[{Chen {et~al.}(2014)Chen, Leung, Boldyrev, Maruca, \& Bale}]{chen14}
Chen, C. H.~K., Leung, L., Boldyrev, S., Maruca, B.~A., \& Bale, S.~D. 2014,
  Geophysical Research Letters, 41, 8081

\bibitem[{Cranmer \& {van Ballegooijen}(2012)}]{cranmer12}
Cranmer, S.~R., \& {van Ballegooijen}, A.~A. 2012, The Astrophysical Journal,
  754, 92

\bibitem[{Fox {et~al.}(2016)Fox, Velli, Bale, Decker, Driesman, Howard, Kasper,
  Kinnison, Kusterer, Lario, Lockwood, McComas, Raouafi, \& Szabo}]{fox16}
Fox, N.~J., {et~al.} 2016, Space Science Reviews, 204, 7

\bibitem[{Horbury {et~al.}(2008)Horbury, Forman, \& Oughton}]{horbury08}
Horbury, T.~S., Forman, M., \& Oughton, S. 2008, Physical Review Letters, 101,
  175005

\bibitem[{Klein {et~al.}(2014)Klein, Howes, \& TenBarge}]{klein14}
Klein, K.~G., Howes, G.~G., \& TenBarge, J.~M. 2014, The Astrophysical Journal
  Letters, 790, L20

\bibitem[{Klein {et~al.}(2015)Klein, Perez, Verscharen, Mallet, \&
  Chandran}]{klein15}
Klein, K.~G., Perez, J.~C., Verscharen, D., Mallet, A., \& Chandran, B. D.~G.
  2015, The Astrophysical Journal Letters, 801, L18

\bibitem[{Kraichnan(1964)}]{kraichnan64}
Kraichnan, R.~H. 1964, Physics of Fluids, 7, 1723

\bibitem[{Lugones {et~al.}(2016)Lugones, Dmitruk, Mininni, Wan, \&
  Matthaeus}]{lugones16}
Lugones, R., Dmitruk, P., Mininni, P.~D., Wan, M., \& Matthaeus, W.~H. 2016,
  Physics of Plasmas, 23, 112304

\bibitem[{Matthaeus {et~al.}(2010)Matthaeus, Dasso, Weygand, Kivelson, \&
  Osman}]{matthaeus10}
Matthaeus, W.~H., Dasso, S., Weygand, J.~M., Kivelson, M.~G., \& Osman, K.~T.
  2010, The Astrophysical Journal Letters, 721, L10

\bibitem[{Matthaeus {et~al.}(2016)Matthaeus, Weygand, \& Dasso}]{matthaeus16}
Matthaeus, W.~H., Weygand, J.~M., \& Dasso, S. 2016, Physical Review Letters,
  116, 245101

\bibitem[{Narita(2017)}]{narita17}
Narita, Y. 2017, Nonlin. Processes Geophys., 24, 203

\bibitem[{Narita {et~al.}(2013)Narita, Glassmeier, Motschmann, \&
  Wilczek}]{narita13}
Narita, Y., Glassmeier, K.-H., Motschmann, U., \& Wilczek, M. 2013, Earth,
  Planets, and Space, 65, e5

\bibitem[{Perez \& Chandran(2013)}]{perez13}
Perez, J.~C., \& Chandran, B. D.~G. 2013, The Astrophysical Journal, 776, 124

\bibitem[{Servidio {et~al.}(2011)Servidio, Carbone, Dmitruk, \&
  Matthaeus}]{servidio11}
Servidio, S., Carbone, V., Dmitruk, P., \& Matthaeus, W.~H. 2011, EPL
  (Europhysics Letters), 96, 55003

\bibitem[{Taylor(1938)}]{taylor38}
Taylor, G.~I. 1938, Proceedings of the Royal Society of London Series A, 164,
  476

\bibitem[{Weygand {et~al.}(2013)Weygand, Matthaeus, Kivelson, \&
  Dasso}]{weygand13}
Weygand, J.~M., Matthaeus, W.~H., Kivelson, M.~G., \& Dasso, S. 2013, Journal
  of Geophysical Research (Space Physics), 118, 3995

\end{thebibliography}

\end{document}